\documentclass[preprint,prl,amsmath]{revtex4}
\usepackage{graphicx}
\begin{document}
\draft
  
\title{Twist and teleportation analogy of the black hole final state}
  
\author{Doyeol Ahn\thanks{Also with Department of Electrical
    Engineering, University of Seoul, Seoul 130-743,
    Korea}\thanks{E-mail; dahn@uoscc.uos.ac.kr}
    and M. S. Kim $^\dagger$
}
\address{Institute of Quantum Information Processing and Systems,
  University of Seoul, \\
  90 Jeonnong, Tongdaemonn-ku, Seoul, Korea\\
E-mail: dahn@uos.ac.kr; davidahn@hitel.net\\
 $^\dagger$ School of Mathematics and Physics, The Queen's University,
  Belfast BT7 1NN, United Kingdom\\
 }

\date{\today}
  
\begin{abstract}
Mathematical connection between the quantum teleportation, the most unique feature of quantum information processing, and the black hole final state is studied taking into account the non trivial spacetime geometry.
We use the twist operatation for the generalized entanglement measurement and the final state boundary conditions
to obtain transfer theorems for the black hole evaporation. This would enable us to put together the universal quantum teleportation
and the black hole evaporation in the unified mathematical footing. 
For a renormalized post selected final state of outgoing Hawking radiation, we found that the measure of 
mixedness is preserved only in the special case of final-state boundary condition in the micro-canonical form, which resmebles perfect teleportation channel.
\end{abstract}


\maketitle

\pagebreak

\section{I. Introduction}
Black hole information paradox \cite{hawk1,hawk2,hawk3} has been a serious challenge to fundamental physics for over three decades. Hawking's semi-classical argument predicts that a process of black hole formation and evaporation is not unitary \cite{hawk3}, which contradicts the basic principles of quantum mechanics-\!- \, evolution of pure state into pure state. On the other hand, there is some evidence in string theory that the formation and evaporation of black hole is a unitary process \cite{HM}. Nonetheless, Hawking effect, discovered nearly 30 years ago, is generally accepted very credible and considered as would be an essential ingredient of the yet unknown correct theory of quantum gravity. 

Previously, Horowitz and Maldacena (HM) proposed a final-state boundary condition \cite{HM} to reconcile the unitarity of the black hole evaporation with Hawking' semi-classical reasoning.  The essence of HM proposal is to impose a unique final boundary condition at the black hole singularity such that no information is absorbed by the singularity. The final boundary state is maximally entangled states of the collapsing matter and infalling Hawking radiation. The projection of final boundary state at the black hole singularity collapses the state into one associated with the collapsing matter and transfer the information to the outgoing Hawking radiation. The HM model is further refined, by including the unitary interactions between the collapsing matter and infalling Hawking radiation \cite{Preskill}, and a random purification of the final boundary state \cite{Lloyd1}. It is also found that the evaporation depends on the boundary condition at the event horizon \cite{Ahn0}.

One of the critical assumptions in the 
HM proposal is that the internal quantum state of the black hole can be represented by maximally entangled states of collapsing matter 
and infalling Hawking radiation. This ansatz is important because the final state boundary condition of the HM proposal is based on this maximally 
entangled internal quantum state \cite{HM}. Recently, one of us proved the HM ansatz for the special case of collapsing gravitational shell 
of the Schwarzschild black hole \cite{Ahn}. 

In the HM proposal, the final state boundary condition (FBC) at the black hole singularity 
acts like a measurement that collapses the state into one associated with an infalling matter \cite{HM}. 
This transfer of the information from the collapsing matter to the outgoing Hawking radiation
is similar to the quantum teleportation \cite{Bennett,Braunstein} as mentioned in the previous works \cite{HM, Preskill}. Quantum teleportation is a unique feature of quantum information processing, which allows moving quantum states around even in the absence of a quantum communication channel, linking the sender of the quantum state to the recipient \cite{Nielsen}. 
Since black hole information paradox is closely related to the aspects of quantum information, it would be an interesting study to relate the black hole evaporation to a quantum teleportation.

In this paper, we compare the joint measurements by local unitary operations required 
for the universal teleportation with the black hole final state and state transfer theorems for the black hole evaporation. Twist operation for the 
generalized entanglement measurement is employed as a principal mathematical tool. 
This would enable us to put together the universal quantum teleportation
and the black hole evaporation in the unified mathematical footing. We first consider the case of a micro-canonical form \cite{HM} of black hole state, which is a limiting case of a more general case such as a Schwarzschild black hole, then we extend the analysis to the case of a Schwarzschild black hole. 

\section{II. Twister and Quantum Teleportation}

We start by reviewing a twister formalism \cite{Braunstein} for the universal teleportation scheme and apply it to 
the black hole evaporation process. Let $\{|n\rangle \}$ be the basis of the $N$ dimensional Hilbert space $H$, $|\ldots \rangle$ be a state in $H$ and $|\ldots \gg$ denote an entangled state in $H \otimes H$. We define the transfer operator as
\begin{equation}
\tau_{b,a}=\sum_{n} |n\rangle_{b} {}_{a}\langle n|,
\end{equation}
the entangled state in $H_{1} \otimes H_{2}$ as
\begin{equation}
|\overrightarrow{\Psi}\gg_{1,2}=\frac{1}{\sqrt{N}} \sum_{n}e^{-i\phi_{n}}|n\rangle_{1} \otimes |n\rangle_{2},
\end{equation}
and its twisted version as
\begin{equation}
|\overleftarrow{\Psi}\gg_{1,2}=\frac{1}{\sqrt{N}} \sum_{n}e^{-i\phi_{n}}|n\rangle_{2} \otimes |n\rangle_{1}.
\end{equation}
Changing the order of states in the tensor products as in eqs. (2) and (3) is called the twist operation \cite{Braunstein}.
In this definition, we would like to note that $|\Psi \gg = |\overrightarrow{\Psi} \gg$.
The twist operation swaps the pair of particles in $H \otimes H$  and we have the following identites
which will be used to describe both teleportation and black hole evaporation process \cite{Braunstein}:
\begin{equation}
{}_{1,2}\ll \overrightarrow{\Psi}|\overleftarrow{\Psi} \gg_{2,3}=\frac{1}{N} \tau_{3,1},
\end{equation}
\begin{equation}
{}_{1,2}\ll \overleftarrow{\Psi}|\overrightarrow{\Psi} \gg_{2,3}=\frac{1}{N} \tau_{3,1},
\end{equation}
and
\begin{equation}
{}_{1,2}\ll \overrightarrow{\Psi}|[ |\phi\rangle_{1} \otimes |\overleftarrow{\Psi} \gg_{2,3}]=\frac{1}{N} |\phi \rangle_{3}.
\end{equation}

Let us use these identities to describe the teleportation process first \cite{Braunstein}. We assume that "2" denotes Alice's state, "3" denotes Bob's state 
and "1" denotes an unknown state.  We start with an unknown state and a maximally entangled resource 
$|\phi\rangle_{1} \otimes |\overrightarrow {\Phi} \gg_{2,3}~$. Alice performs a generalized measurement on $H_{1} \otimes H_{2}~$
and obtains a maximally entangled result $|\overrightarrow{\Psi}\gg_{1,2}$. Alice then let Bob know her measurement result.
To accomplish the teleportation procedure, Bob must now convert $|\overrightarrow {\Phi} \gg_{2,3}$ into the twisted version of
the entangled state found by Alice $|\overleftarrow {\Psi} \gg_{2,3}$ after the generalized or overcomplete measurment.
Then from eq. (6), the initial unknown state $|\phi \rangle_{1}$ is teleported to Bob.

The set of all maximally entangled states can be represented by local unitary operations \cite{Braunstein}. 
We denote the unitary operator as $S=S(g)$ where $g$ is an element a group $G=\{ g \}$ of transformations.
For example, 
\begin{equation}
|\overrightarrow {\Psi} \gg_{1,2}=S(g) \otimes I |\overrightarrow {\Phi} \gg_{1,2}
\end{equation}
for particular group operation $g$.
Once Alice lets Bob know the unitary operation $S(g)$ over classical channel, then Bob 
can perform twist operation on $|\overrightarrow {\Phi} \gg_{1,2}$ to complete the teleporation.

\section{III. Final sate in micro-canonical form}

It was pointed out in the previous works \cite{HM,Preskill,Lloyd1,Ahn0,Ahn} that the final state boundary condition resembles 
generalized measurement for the quantum teleportation. However, the resemblance ends here because during the black hole
evaporation there is no way to communicate the information of $S(g)$ through the event horizon to outside of the black hole. In the following, 
we reformulate the HM proposal using the twist operation developed in the previous section and extend the
proposal to the case of a mixed state.
In the HM model, the boundary state outside the event horizon is assumed to be the Unruh vacuum state \cite{Unruh, Wald, Davies}. 


We assume that the quantum state of the collapsing matter belongs to a Hilbert space $H_{M}$ with dimension $N$  and $|\phi\rangle_{M}~$ be the initial quantum state of the collapsing matter. 
The Hilbert space of fluctuations on the background spacetime for black hole formation and evaporation 
is separated into $H_{in}$  and $H_{out}$  which contain quantum states localized inside and outside the event horizon, respectively. 
The Unruh vacuum state $|\Phi\rangle_{in\otimes out}$   belonging to $H_{in}\otimes H_{out}~$  
in micro-canonical form is given by \cite{HM, Preskill, Lloyd1, Ahn0,Ahn}:

\begin{equation}
|\Phi\gg_{in,out}=|\Phi \rangle_{in\otimes out}={1\over\sqrt{N}}\sum_{l=1}^N|{l}\rangle_{in}\otimes|{l}\rangle_{out}~,
\end{equation}
where $\{|{l}\rangle_{in}\}$  and  $\{|{l}\rangle_{out}\}$   are orthonormal bases for  $H_{in}$ and $H_{out}$ , respectively. 
The final-state boundary condition (FBC) imposed at the singularity requires 
a maximally entangled quantum state in  $H_{M}\otimes H_{in}$ which is called final boundary state and is given by
\begin{equation}
{}_{M,in}\ll\overrightarrow{\Psi}|={}_{M\otimes in}\langle \Psi|={1\over \sqrt{N}}\sum_{l=1}^N {}_M\langle {l}|\otimes {}_{in}\langle {l} |(S^{\dagger}(g) \otimes I)~,
\end{equation}
where $S(g)$ is a unitary transformation. 

In HM proposal the evaporation process is defined as the generalized measurement 
on the state $|\phi\rangle_{M} \otimes |\Phi \gg_{in,out}$ by the black hole final-state $|\overrightarrow{\Psi} \gg_{M,in}~$. From this, we state the following conjecture.

\emph{\bf Horowitz-Maldacena conjecture:} \ \emph{Given states} $|\phi\rangle_{M} \otimes |\Phi \gg_{in,out}$
\emph{and} $|\overrightarrow{\Psi} \gg_{M,in}$, \emph{the evaporation process which transforms initial collapsing matter to the state of outgoing Hawking radiation is described by the final state projection}
${}_{M,in}\ll \overrightarrow{\Psi}|[ |\phi\rangle_{M} \otimes |\Phi \gg_{in,out}]$. 

We first obtain following trnasfer operator
\begin{eqnarray}
\Omega_g & = & {}_{M,in} \ll \overrightarrow{\Psi}|\Phi \gg_{in,out} \nonumber \\
& = & \frac{1}{N}\sum_{k,l}({}_{M}\langle k|S^{\dagger}(g)\otimes {}_{in}\langle k|)(|l \rangle_{in} \otimes |l \rangle_{out}) \nonumber \\
& = & \frac{1}{N} \sum_{k} |k \rangle_{out} {}_{M} \langle k| S^{\dagger}(g) \nonumber \\
& = & \frac{1}{N} \tau_{out,M} S^{\dagger}(g). 
\end{eqnarray}
Eq. (10) implies the following transfer theorem \cite{HM}:

{\em \bf Theorem \ 1:} \ \emph{Given state} $|\phi\rangle_{M} \otimes |\Phi \gg_{in,out}$,\ \emph{the black hole final state} $|\overrightarrow{\Psi} \gg_{M,in}$, \emph{and the transfer operator} $\Omega_g$, \emph{the final state projection yields} 
\begin{eqnarray}
{}_{M,in}\ll \overrightarrow{\Psi}|[ |\phi\rangle_{M} \otimes |\Phi \gg_{in,out}]  
& = & \Omega_g |\phi \rangle_{M} \nonumber \\
& = & \frac {S^{\dagger}(g)}{N}|\phi \rangle_{out}.
\end{eqnarray}
The proof of the transfer theorem is as follows:
\begin{eqnarray}
&   &{}_{M,in}\ll \overrightarrow{\Psi}|[ |\phi\rangle_{M} \otimes |\Phi \gg_{in,out}]  \nonumber \\
& = &\frac{1}{N}\sum_{k,l}{}_M\langle k|S^{\dagger}(g)\otimes{}_{in}\langle k|[|\phi \rangle_M \otimes (|l \rangle_{in} \otimes |l \rangle_{out})] \nonumber \\
& = &\frac{1}{N}\sum_{k}|k \rangle_{out} {}_{M}\langle k|S^{\dagger}(g)|\phi \rangle_{M} \nonumber \\
& = &\frac{1}{N}\sum_{k}|k \rangle_{out} \langle k|S^{\dagger}(g)|\phi \rangle_{out} \nonumber \\
& = & \frac {S^{\dagger}(g)}{N}|\phi \rangle_{out}.
\end{eqnarray}
From eqs. (10) to (12), we can see that the twister with a transfer operator simplies the calculation significantly.  
The initial matter state $|\phi\rangle_{M}$ evolves into the state of outgoing Hawking radiation 
$\frac {S^{\dagger}(g)}{N}|\phi \rangle_{out}$. This clearly indicates that the pure state evolves
into the pure state during the evaporation process.

Then, one may ask the following questions:\emph{"Does quantum mechanics dictate the mixed state evolves into the mixed state?"} and \emph{"Is the measure of mixedness preserved at evaporation?"}

Mixed state is represented 
by the density operator of the form,
\begin{equation}
\rho_{M}=\sum_{n}C_n |n\rangle_{M}\langle n|,
\end{equation}
with
\begin{equation}
\sum_{n}C_n=1.
\end{equation}
Here we have used the fact that the density operator is a positive operator.
A mixed state arises when the initial matter state was entangled with other quantum systems 
when the black hole was about to be formed and was taken a partial trace over these other systems.
The measure of mixedness is defined by 
\begin{equation}
M_x=Tr(\rho^2_M).
\end{equation}

In order to get the final density operator at evaporation, we need to calculate
${}_{M,in}\ll~\overrightarrow{\Psi}|[\rho_{M} \otimes |\Phi \gg_{in,out} \ll\Phi|]|\overleftarrow{\Psi} \gg_{M,in}~$
to study the evolution of mixed state.
We first obtain the following identities:
\begin{equation}
{}_{M,in}\ll \overrightarrow{\Psi}|\Phi \gg_{in,out}=\frac{1}{N} \tau_{out,M}S^{\dagger}(g),
\end{equation}
and
\begin{equation}
{}_{in,out}\ll \Phi|\overleftarrow{\Psi} \gg_{M,in}=\frac{S(g)}{N} \tau_{M,out}.
\end{equation}
From these identities, we state the transfer theorem for the mixed state.

\emph{\bf Theorem \ 2:} \ \emph{Given states} $\rho_{M} \otimes |\Phi \gg_{in,out} \ll\Phi|$ \emph{and}
$|\overleftarrow{\Psi} \gg_{M,in}$, \emph{the final state projection yields}
\begin{eqnarray}
&   &{}_{M,in}\ll \overrightarrow{\Psi}|[\rho_{M} \otimes |\Phi \gg_{in,out} \ll\Phi|]|\overleftarrow{\Psi} \gg_{M,in} \nonumber \\
& = &\frac{S^{\dagger}(g)}{N^2}\tau_{out,M}\rho_M\tau_{M,out}S(g) \nonumber \\
& = &\frac{1}{N^2}S^{\dagger}(g)(\sum_{n} C_n|n \rangle_{out} \langle n|)S(g) \nonumber \\
& = & \frac{1}{N^2} S^{\dagger}(g) \rho_{out}S(g).
\end{eqnarray}
In order to prove this, we first calculate
\begin{eqnarray}
&   &\rho_M \otimes |\Phi \gg_{M,out}\ll\Phi| \nonumber \\
& = &\frac{1}{N}\sum_{k,l,n}C_n|n \rangle_{M}\langle n| \otimes (|k\rangle_{in}\langle l|)\otimes (|k\rangle_{out}\langle l|). 
\end{eqnarray}
Then,
\begin{eqnarray}
&   &{}_{M,in}\ll \overrightarrow{\Psi}|[\rho_{M} \otimes |\Phi \gg_{in,out} \ll\Phi|]|\overleftarrow{\Psi} \gg_{M,in} \nonumber \\
& = & \frac{1}{N^2}\sum_{k,l,m,n,p}C_n \ ({}_M \langle m|S^{\dagger}(g) \otimes {}_{in} \langle m|) \otimes (|n \rangle_{M} \langle n|) \otimes (|k \rangle_{in} \langle l|) \nonumber \\
&   & \otimes (|k \rangle_{out} \langle l|) S(g) |p \rangle_{M} \otimes |p \rangle_{in} \nonumber \\
& = & \frac{1}{N^2} \sum_{m,n,p} C_n \ {}_M \langle m|S^{\dagger}(g)|n \rangle_M \langle n|S(g)|p \rangle_M \otimes |m \rangle_{out} \langle p| \nonumber \\
& = & \frac{1}{N^2} \sum_{m,n,p} C_n|m \rangle_{out} \langle m|S^{\dagger}(g)|n \rangle_{out} \langle n|S(g)|p \rangle_{out}   \langle p| \nonumber \\
& = & \frac{1}{N^2}\sum_{n}C_n \ S^{\dagger}(g)|n \rangle_{out} \langle n|S(g) \nonumber \\
& = & \frac{1}{N^2} S^{\dagger}(g) \rho_{out} S(g).  
\end{eqnarray}
Here we have used the closure relation $\sum_n |n \rangle \langle n|=1.$
We can see that the initial mixed state evolves into a mixed state. 
In the final state boundary condition, the evaporation is described by the
unitary representation $S(g)$ of the group element $g$ of $G$.
The normalization factor $\frac{1}{N^2}$ indicates that the outcome of the generalized measurement
representing the evaporation would occur with probability $\frac{1}{N^2}$. 
Then eq. (20) suggests that the final density operator is "post selected" state \cite{HM}. 
All the other outcomes are to be discarded. By following the generalized measurement process,
one can renormalize the post selected state\cite{Nielsen}, which is given by
\begin{equation}
\rho_f=S^{\dagger}(g) \ \rho_{out} S(g),
\end{equation}
where
\begin{equation}
\rho_{out}=\sum_{n} C_n \ |n \rangle_{out} \langle n|.
\end{equation}

We now study how the mixedness changes under evaporation. The measure of mixedness of the initial matter state
is defined by
\begin{eqnarray}
M_{xi}& = &Tr(\rho^2_M) \nonumber \\
& = & Tr(\sum_{n,m}C_n \ C^{\ast}_m |n \rangle \langle n| \ |m \rangle \langle m|) \nonumber \\
& = & Tr (\sum_n |C_n|^2 \ |n \rangle \langle n|) \nonumber \\
& = & \sum_n |C_n|^2 \ < \ 1.
\end{eqnarray}
The measure of mixedness of the final density operator is given by
\begin{eqnarray}
M_{xf}& = &Tr(\rho^2_f) \nonumber \\
& = & Tr(S^{\dagger}(g) \rho_{out} \ S(g) \ S^{\dagger} (g) \ \rho_{out} \ S(g))) \nonumber \\
& = & Tr (\rho^2_{out}) \nonumber \\
& = & M_{xi}.
\end{eqnarray}
This immediately suggests that the measure of mixedness is preserved during the black hole evaporation in the case of the final-state boundary condition in micro-canonical form.

\section{IV. Final state for the Schwarzschild black hole}
Representation of Hawking radiation by the micro-canonical form considered in the previous section can be considered as a limiting case of the Bogoliubov transformed vacuum for the Kruskal-Schwarzschild spacetime assocaited with a Schwarzschild black hole. TheHawking radiation of the Schwarzschild
black hole is described by the Bololiubov transformed vacuum which yields infinite dimensional two-mode squeezed state \cite{Ahn}. This situation is analogous to the imperfect teleportation channel \cite{Braunstein}. 

The Unruh vacuum state $|\Phi(\lambda) \gg_{in,out}$ belonging to $H_{in} \otimes H_{out}~$ is given by \cite{Ahn}
\begin{equation}
|\Phi(\lambda) \gg_{in,out} = (1-\lambda^{2})^{1/2} \sum_{n} \lambda^{n} |n \rangle_{in} \otimes |n \rangle_{out},
\end{equation}
where 
$\lambda$ is a physical parameter describing the Hawking radiation, $\lambda \in \, [0,1)$
such that the state becomes micro-canonical form as $\lambda \rightarrow 1$. Crucial aspects of black hole physics related to the non trivial spacetime geometry is contained in $\lambda$ \cite{Ahn, Wald3, Parker}.
In the case of the Schwarzschild black hole, $\lambda=e^{-4\pi M \omega}~$, where $M$ is the mass of the black hole and $\omega$ is the positive frequency of the normal mode.
As a black hole internal state, we have \cite{Ahn, Fabbri}
\begin{equation}
|\Phi(\lambda) \gg_{M,in} = (1-\lambda^{2})^{1/2} \sum_{n} \lambda^{n} |n \rangle_{M} \otimes |n \rangle_{in}.
\end{equation}
The final state boundary condition is then given by \cite{Ahn}
\begin{equation}
|\overrightarrow{\Psi(\lambda)} \gg_{M,in} =(S(g) \otimes I) |\Phi(\lambda) \gg_{M,in}.
\end {equation}
Now, we state the following two lemmas.

\emph{\bf Lemma 1:}\, \emph{Given states} \, $|\Phi(\lambda) \gg_{M,in}$ \emph{and}
$|\overrightarrow{\Phi(\lambda)} \gg_{M,in}$, \emph{we obtain}
\begin{equation}
{}_{M,in} \ll \overrightarrow{\Psi(\lambda)}|\Phi(\lambda) \gg_{in,out}=D(\lambda)S^{\dagger}(g)\tau_{out,M},
\end{equation}
\emph{where the distortion operator} $D(\lambda)$ is defined by
\begin{equation}
D(\lambda)=(1-\lambda^2) \sum_{n} \lambda^{2n} |n \rangle \langle n|.
\end{equation}
\emph{Further more}, $Tr(D(\lambda))=1$ \emph{and} $Tr(D^2(\lambda) \leq 1$.

\emph{\bf Lemma 2:}\, \emph{Given states} \, $|\Phi(\lambda) \gg_{M,in}$ \emph{and}
$|\overrightarrow{\Phi(\lambda)} \gg_{M.in}$, \emph{we obtain}
\begin{equation}
{}_{in,out} \ll \Phi(\lambda)|\overleftarrow{\Psi(\lambda)} \gg_{in,out}=\tau_{M,out}S(g)D(\lambda).
\end{equation}

The proof of the lemma 1 is as follows:
\begin{eqnarray}
{}_{M,in} \ll \overrightarrow{\Psi(\lambda)}|\Phi(\lambda) \gg_{in,out} 
& = & (1-\lambda^2) \sum_{n,m} \lambda^{n+m}({}_M \langle n|S^{\dagger}(g) \otimes {}_{in} \langle n|)(|m \rangle_{in} \otimes |m \rangle_{out} \nonumber \\
& = & (1-\lambda^2) \sum_{n} \lambda^{2n} |n \rangle_{out} \, {}_{M} \langle n| S^{\dagger}(g) \nonumber \\
& = & (1-\lambda^2) \sum_{n,m} \lambda^{2n} |n \rangle_{out} \, {}_{M} \langle n| S^{\dagger}(g) |m \rangle_{M} \langle m| \nonumber \\
& = & (1-\lambda^2) \sum_{n,m} \lambda^{2n} |n \rangle_{out} \langle n| S^{\dagger}(g) |m \rangle_{out} \, {}_{M} \langle m| \nonumber \\
& = &D(\lambda)S^{\dagger}(g)\tau_{out,M}
\end{eqnarray}
The proof of lemma 2 is straightforward and can be obtained by the same method. 
The following two theorems stated here without proof, describe the evaporation of
pure state and mixed state in the case of final-state boundary condition for the Schwarzschild black hole. 

{\em \bf Theorem \ 3:} \ \emph{Given state} $|\phi\rangle_{M} \otimes |\Phi(\lambda) \gg_{in,out}$,\ \emph{the black hole final state projection yields}
\begin{equation}
{}_{M,in} \ll \overrightarrow{\Psi(\lambda)}| [|\phi \rangle_{M} \otimes | \Phi(\lambda) \gg_{in,out}]=D(\lambda)S^{\dagger}| \phi \rangle_{out}.
\end{equation}

\emph{\bf Theorem \ 4:} \ \emph{Given states} $\rho_{M} \otimes |\Phi(\lambda) \gg_{in,out} \ll\Phi(\lambda)|$ \emph{and}
$|\overleftarrow{\Psi(\lambda)} \gg_{M,in}$, \emph{the black hole final state projection yields}
\begin{eqnarray}
&   &{}_{M,in}\ll \overrightarrow{\Psi(\lambda)}|[\rho_{M} \otimes |\Phi(\lambda) \gg_{in,out} \ll\Phi(\lambda)|]|\overleftarrow{\Psi(\lambda)} \gg_{M,in} \nonumber \\
& = &D(\lambda)S^{\dagger}(g)\tau_{out,M}\rho_M\tau_{M,out}S(g)D(\lambda) \nonumber \\
& = &D(\lambda) S^{\dagger}(g) \rho_{out}S(g)D(\lambda).
\end{eqnarray}
The proofs of theorem 3 and theorem 4 are straightforward.

Eqs. (32) and (33) suggest that the HM proposal can still be extended to the general case. A pure state evoles into a pure state and a mixed state evolves into a mixted state. 
Comparing eqs. (20) and (33), we note that the black hole final state projection
looks like the nonlocal quantum operation $\rho_{M} \rightarrow \mathcal{E}(\rho_{out})=E(g)\rho_{out}E^{\dagger}(g)$, where $E(g)$ is given by
\begin{eqnarray}
E(g)=\left \{ \begin{array}{c}
\frac{S^{\dagger}}{N} \, \, \, ,\, \mbox{for FBC in micro-canonical form,}   \\
D(\lambda)S^{\dagger}(g) \, \, ,\, \mbox{for FBC for Schwarzschild black hole.}
\end{array}
\right .
\end{eqnarray} 
In the case of FBC in micro-canonical form, $E^{\dagger}(g)E(g)=\frac{S(g)}{N} \frac{S^{\dagger}(g)}{N}=\frac{1}{N^2}~$
and if the number of group element is $N^2$, then we have $\sum_{g}E^{\dagger}(g)E(g)=1$. This imples that
$E(g)$ describes the general measurement \cite{Nielsen, Kraus} and the group element $g$ specifies a particular measurement. 
On the other hand, in the case of FBC for a Schwarzschild black hole,
we get $E^{\dagger}(g)E(g)=S(g)D^2(\lambda)S^{\dagger}(g)$ with $tr(E^{\dagger}(g)E(g))=Tr(D^2(\lambda) \leq 1$.
Here $D(\lambda)$ resembles the Kraus operator for decoherence process \cite{DAhn}. Under the HM conjecture,
the evaporation can be described by the nonlocal teleportation of $\rho_M$ to $\rho_{out}$ followed by the 
quantum operation $\mathcal{E}(\rho_{out})$. In this case, the quantum operation $\mathcal{E}$ is 
non-trace preserving $(Tr({E}(\rho_{out}) \leq 1)$, convex-linear, and completely-positive. Non-trace preseving 
nature of the quantum operation ${E}(\rho_{out})$ means that we do not have the complete description of 
the evaporation process \cite{Nielsen}. For example, we do not know the properties of a quantum group $G$, yet.
One can still renormlize the final density operator to get
\begin{equation}
\rho_f=\frac{D(\lambda)S^{\dagger}(g) \rho_{out}S(g)D(\lambda)}{Tr(D(\lambda)S^{\dagger}(g) \rho_{out}S(g)D(\lambda))}.
\end{equation} 
The measure of mixedness is defined by
\begin{eqnarray}
Tr(\rho^{2}_{f}) & = & \frac{Tr[D(\lambda)S^{\dagger}(g) \rho_{out}S(g)D^2(\lambda)S^{\dagger}(g) \rho_{out}S(g)D(\lambda)]}{Tr^2(D(\lambda)S^{\dagger}(g) \rho_{out}S(g)D(\lambda))} \nonumber \\
& = & \frac{Tr[\rho_{out}S(g)D^2(\lambda)S^{\dagger}(g) \rho_{out}S(g)D^2(\lambda)S^{\dagger}(g)]}{Tr^2(\rho_{out}S(g)D^2(\lambda)S^{\dagger}(g))} \nonumber \\
& = & \frac{Tr[\rho_{out}W_{\lambda}(g) \rho_{out}W_{\lambda}(g)]}{Tr^2(\rho_{out}W_{\lambda}(g))},
\end{eqnarray}
where $W_{\lambda}(g)$ is defined by
\begin{equation}
W_{\lambda}(g)=S(g)D^2(\lambda)S^{\dagger}(g). 
\end{equation}
We define $\tilde{W}_{\lambda}(g)$ by
\begin{equation}
\tilde{W}_{\lambda}(g)=\frac{W_{\lambda}(g)}{Tr(\rho_{out}W_{\lambda}(g))}.
\end{equation}
Then eq. (36) can be rewritten as
\begin{equation}
Tr(\rho^{2}_{f})=Tr[\tilde{W}_{\lambda}(g) \rho_{out} \tilde{W}_{\lambda}(g)\rho_{out}].
\end{equation}
For all bounded linear operator $X$ and density operator $T$, we have the following inequlaity \cite{Kraus}
\begin{equation}
|Tr(XT)|\leq  \parallel X \parallel \, \parallel T \parallel_{1},
\end{equation}
where
$\parallel \cdot \parallel$ is an operator norm and $\parallel \cdot \parallel_{1}$ is a trace norm.
Then, we have
\begin{eqnarray}
Tr(\rho^{2}_{f}) & = & Tr[\tilde{W}_{\lambda}(g) \rho_{out} \tilde{W}_{\lambda}(g)\rho_{out}] \nonumber \\
& \leq & \parallel \tilde{W}_{\lambda}(g) \parallel Tr[\rho_{out} \tilde{W}_{\lambda}(g)\rho_{out}] \nonumber \\
& = & \parallel \tilde{W}_{\lambda}(g) \parallel Tr[\tilde{W}_{\lambda}(g)\rho_{out}^2] \nonumber \\
& \leq & \parallel \tilde{W}_{\lambda}(g) \parallel^2 Tr[\rho_{out}^2] \nonumber \\
& \leq & Tr[\rho_{out}^2],
\end{eqnarray}
since $\sup \tilde{W}_{\lambda}(g)=1$.
This suggests that the measure of mixedness is expected to decrease (or more mixed) under 
evaporation in the case of final state boundary condition with non-uniform distribution. 
On the other hand, in the case of FBC in micro-canonical form, originally proposed by
Horowitz and Maldacena \cite{HM}, we found that the black hole information paradox is resolved perfectly
for both a pure and a mixted state.  In the pursuit of the resolution of the black hole information paradox, we were
concerned with the evolution of the pure state under evaporation until now. We may need a further study
to see how the evolution of a mixed state and especially the preservation of the measure of mixedness are
related to the black hole information paradox in the general case.

\section{V. Summary}
In this paper, we studied unified formulation of 
the quantum teleportation and the black hole evaporation
based on a twist operator for the 
generalized entanglement measurement taking into account the non-trivial spacetime geometry. This enables us to put together the universal quantum teleportation
and black hole evaporation in the unified mathematical footing. 
The resolution of the black hole information paradox 
using the final state boundary condition implies not only the pure state evolving into the pure state
but also the mixed state evolving into the mixed state. 
It is found that for the latter case the measure of 
mixedness is preserved only in the case of final-state boundary condition in micro-canonical form.

\vspace{1cm}

\acknowledgements 

This work was supported by KOSEF and MOST through the Creative Research Initiatives Program R-16-1998-009-01001-0(2006).


\begin{references} 
  
\bibitem{hawk1} S.~W. Hawking, ``Black hole explosions,'' Nature {\bf 248}, 30 (1974).
\bibitem{hawk2} S.~W. Hawking, ``Particle creation by black holes,'' Commun. Math. Phys. {\bf 43}, 199 (1975).
\bibitem{hawk3} S.~W. Hawking, ``Breakdown of predictability in gravitational collapse,'' Phys. Rev. D {\bf 14}, 2460 (1976).
\bibitem{HM} G.~T. Horowitz and J. Maldacena, ``The black hole final state,'' J. High Energy Phys.  {\bf 02}, 008 (2004).
\bibitem{Preskill} D. Gottesman and J. Preskill, ``Comment on "The black hole final state",'' J. High Energy Phys.  {\bf 03}, 026 (2004).
\bibitem{Lloyd1} S. Lloyd, ``Almost certain escape from black holes in final state projection models,'' Phys. Rev. Lett.  {\bf 96}, 061302 (2006).
\bibitem{Ahn0} D. Ahn, ``Control of black hole evaporation?''  [hep-th/0604196].
\bibitem{Ahn} D. Ahn, ``Final state boundary condition of the Schwarzschild black hole,'' Phys. Rev. D  {\bf 74}, 084010 (2006).
\bibitem{Bennett} C.~H.~Bennett, G.~Brassard, C.~Crepeau, R.~Jozsa, A.~Peres and W.~K.~Wootters,
``Teleporting an unknown quantum state via dual classical and Einstein-Podolsky-Rosen channels,''
Phys. Rev. Lett. {\bf 70}, 1895 (1993).
\bibitem{Braunstein} S.~L. Braunstein, G.~M. D'Ariano, G.~J. Milburn and M.~F. Sacchi``Universal teleportation with a twist,'' Phys. Rev. Lett.  {\bf 84}, 3486 (2000).
\bibitem{Nielsen} M.~A. Nielsen and I. L. Chuang, ``Quantum Computation and Quantum Information,'' (Cambridge University Press, New York, 2000).
\bibitem{Unruh} W.~G. Unruh, ``Notes on black-hole evaporation,'' Phys. Rev. D  {\bf 14}, 870 (1976).
\bibitem{Wald} R.~M. Wald, ``Quantum field theory in curved spacetime and black hole thermodynamics,'' (The University of Chicago Press, Chicago, 1994).
\bibitem{Davies} N.~D. Birrell and P.~C.~W. Davies, ``Quantum field theorys in curved space,'' (Cambridge University Press, New York, 1982).
in private communication.
\bibitem{Wald3} R. M. Wald, ``On particle creation by black holes,'' Commun. Math. Phys.  {\bf 43}, 9 (1976).
\bibitem{Parker} L. Parker, ``Probability distributions of particles created by a black hole,'' Phys. Rev. D  {\bf 12}, 1519 (1976).
\bibitem{Fabbri} A. Fabbri and A. Perez, ``Black hole evaporation in thermalized final-state projection model'' [hep-th/0604196].
\bibitem{Kraus} K. Kraus, ``States, Effects, and Operations,'' (Springer-verlag, New York, 1983).
\bibitem{DAhn} D. Ahn, J. Lee, M.~S. Kim, and S.~W. Hwang ``Self-consistent non-Markovian theory of a quantum-state evolution for quantum information processing,'' Phys. Rev. A  {\bf 66}, 012302 (2002).

\end{references}
\end{document}